\documentclass[pra,aps,amsmath,twocolumn,floatfix]{revtex4-1}
\usepackage{amsmath,graphicx}
\begin{document}
\def\tr{\rm{Tr}}
\def\la{{\langle}}
\def\ra{{\rangle}}
\def\a{{\alpha}}
\def\e{\epsilon}
\def\q{\quad}
\def\w{\tilde{W}}
\def\t{\tilde{t}}
\def\a{\hat{A}}
\def\h{\hat{H}}
\def\E{\mathcal{E}}
\def\p{\hat{P}}
\def\u{\hat{U}}
\def\n{\\ \nonumber}
\def\j{\hat{j}}
\def\alph{a}
\def\vc{\underline{c}}
\def\vf{\underline{f}}
\title{Comment on the MS  "Null weak values and the past of quantum particle" by  Q.Duprey and  A.Matzkin}
\author {D. Sokolovski$^{a,b}$}
\affiliation{$^a$ Departmento de Qu\'imica-F\'isica, Universidad del Pa\' is Vasco, UPV/EHU, Leioa, Spain}
\affiliation{$^b$ IKERBASQUE, Basque Foundation for Science, E-48011 Bilbao, Spain}
\date{\today}
\begin{abstract}
In a recent paper \cite{Matz}, Duprey and Matzin investigeated the meaning of vanishing "weak values" (WV), 
and their role in the retrodiction of the past of a pre- and post-selected quantum system in the presence of interference.
Here we argue that any proposition, regarding the WV values, should be understood as a statement about the probability amplitudes, and revisit some of the conclusions reached in \cite{Matz}.
\end{abstract}
\maketitle
\noindent
{Keywords; {\it quantum particle's  past, transition amplitudes, weak measurements}}
\vspace{0.5cm}
\section{Introduction}
In a recent publication \cite{Matz}, Duprey and Matzin analysed physical significance of the vanishing weak values.
They also commented on our related work \cite{PLA2017}, suggesting that its approach " discards any possibility to infer a property from protocols implementing nondestructive weak interactions." Here we clarify our position on the issue, 
and discuss some of the conclusions arrived at  in  \cite{Matz}. We start with the basics.

Standard quantum mechanics {\it postulates} the existence of probability amplitudes, whose evaluation inevitably precedes the calculation of probabilities, related to the frequencies with which physical phenomena are observed. 
The amplitudes are complex valued quantities with no specific prescription for the signs of their real and imaginary parts.
The rules for using them are well known (see, for example, the three points of the Summary on the page 1-10 of \cite{Feynl}).
Any "deeper meaning" these axiomatic quantities might have is not available to us. Feynman's statement that "we have no ideas about a more basic mechanism..." \cite{Feynl} remains, we argue, valid to this day.
\section{Virtual pathways and functionals}
As the first point, we note next that our analysis need not be limited to quantities represented by hermitian operators, 
so that the "eigenstate-eigenvalue link" often mentioned in  \cite{Matz} is of no particular importance to us.
Consider a system going from  $|\psi_I\ra$ at $t_1$  to  $|\psi_F\ra$ at $t_2$ via $N$ alternative ways (pathways, paths),  $\{i\}$, $i=1,2,..,N$, each endowed with the probability amplitude $A^{F\gets I}_i$.
To distinguish between the pathways, we may employ a functional $F$, taking a real value
$\mathcal{F}_i$ on the $i$-th path,
\begin{eqnarray}\label{1}
F[i]=\sum_{i} \mathcal{F}_jF^j[i] , \q  F^j[i] =\delta_{ij},
\end{eqnarray} 
where $\delta_{ij}$ is the Kronecker delta.
As a simple example, consider  an $F$ representing the difference between the values of an operator $\hat B$
at some intermediate times $t'$ and $t''$. In a two dimensional Hilbert space $\mathcal{H}$, and for a $\hat B$ with eigenvalues
of $\pm1$, the said difference can take three distinct values of $2$, $0$, and $-2$. It is clear that a measurement of $F$ cannot be reduced to projecting onto the eigenstates of an operator, since no operator acting in $\mathcal{H}$
can have more than two different eigenvalues \cite{DSMath}. Other examples of functionals referring to more than one moment in time, include the residence time of a qubit \cite{QRES}, 
time average of a dynamical variable \cite{TIMAV},
and the quantum traversal time \cite{QTT}.
\section{Accurate (strong) measurements with post-selection}
In can be shown \cite{DSMath} that  if  $K\le N$ of the $\mathcal{F}$s are different, an {\it accurate} (strong) meter would destroy interference between the unions (superpositions \cite{DSMath}) of paths 
corresponding to the different values of $F$.
It will, therefore, create $K$ exclusive routes \cite{Feynl} 
endowed with {\it both} the amplitudes [$\Delta(a-b)=1$ for $a=b$, and $0$ otherwise]
\begin{eqnarray}\label{2}
\tilde{A}^{F\gets I}_k = \sum_{i=1}^N A^{F\gets I}_i\Delta(\mathcal{F}_i-\mathcal{F}_k),\q k=1,2,...,K,
\end{eqnarray}
{\it and} the probabilities $P^{F\gets  I}_k=|\tilde {A}^{F\gets I}_k|^2$
An accurate pointer would always point at one of the $\mathcal{F}_k$, so that  for  its mean shift we have
\begin{eqnarray}\label{3}
\la f\ra_s = \sum_{i=1}^K  \mathcal{F}_k P^{F \gets  I}_k/\sum_{j=1}^KP^{F\gets I}_j.
\end{eqnarray} 
Importantly, Eq.(\ref{2}) is a statement about the probabilities $P^{F \gets  I}_k$, 
{\it created} by our accurate meter. Namely, we learn that multiplying them by $ \mathcal{F}_k$, and adding up  the products, would yield the number in the l.h.s. of Eq.(\ref{2}). 
If we can accurately measure a functional $F_M[i]$  taking the value of $1$ on, say, the first  $M$ paths and $0$ on the rest of them, 
Eq.(\ref{3}), with $K=2$, will yield the probability for travelling the exclusive (real) route, given by superposition of  the said $M$ paths.
\section{Inaccurate (weak) measurements with post-selection}
On the  other hand, an {\it inaccurate} (weak) meter would perturb the system only slightly, and {\it not} create probabilities for individual pathways, or their unions.
The "weakness" can be achieved by either reducing the coupling to the meter, or by broadening the initial pointer state in the coordinate space (see Sect. 10 of \cite{PLA2015}). It is easy to show \cite{Ah1}, \cite{DSMath} that the mean  pointer shift may be given by
\begin{eqnarray}\label{4}
\la f\ra_w = \text{Re} (\text{Im})  \left [\sum_k  \mathcal{F}_k \tilde{\alpha}^{F \gets  I}_k\right ],\q 
\tilde{\alpha}^{F  \gets  I}_k \equiv \frac{\tilde{A}^{F \gets  I}_k}{\sum _{j=1}^K \tilde{A}^{F \gets  I}_j}, \q
\end{eqnarray}
where the choice of the real or imaginary part depends on how the measurement is set up \cite{Ah1}, \cite{DSMath}.
In the absence of probabilities, Eq.(\ref{4}) is a statement about the (relative) probability amplitudes $\tilde{\alpha}^{F  \gets  I}_k $.
Multiplying them by $ \mathcal{F}_k$, adding them up, and taking the real (imaginary) parts, would yield the number in the l.h.s. of Eq.(\ref{4}).
A weak measurement of  the $F_M[i]$, mentioned above, will yield the amplitude for the pathway uniting the first  $M$ paths.
\section{Null weak value of a projector}
The general principle just outlined, applies also to the particular case studied  in \cite{Matz}.
Now the quantity of interest is the instantaneous value of an operator $\hat{B}=\sum_{i=1}^N |b_i\ra B_i \la b_i|$, with the eigenstates 
$|b_i\ra$ and eigenvalues $B_i$, evaluated at some $t_1 < t<t_2$. In general, $K\le N$ of the eigenvalues $B_i$ may be different. We, therefore, have
\begin{eqnarray}\label{5}
.\mathcal{F}_k=B_k, \q \text{and}\q \q\q\q\q\q\q\n
 \tilde{A}^{F\gets I}_k = \sum_{i=1}^N \la \psi_F| \hat{U}(t_2,t)|b_i\ra \la b_i|\hat{U}(t,t_1)|\psi_I\ra 
 \Delta(B_i-B_k),
\end{eqnarray}
where $\hat{U}(t_2,t)$ is the system's evolution operator. 
\newline
Following Ref. \cite{Matz}, we first look at a "null weak value of a projecting operator", $K=2$, and $B_k=\delta_{km}$. In agreement with the above, the authors of \cite{Matz} note that a weak measurement 
of this projector "picks up a relative path amplitude", $\tilde{\alpha}^{F \gets  I}_m=\la \psi_F| \hat{U}(t_2,t)|b_m\ra \la b_m|\hat{U}(t,t_1)|\psi_I\ra/ \la \psi_F| \hat{U}(t_2,t_1)|\psi_I\ra$, which, in this case, happens to be zero.  They proceed, however, to conclude
that "it is meaningless to make any assertion concerning the property of the system if interferences are not lifted by a strong coupling". This is not our position.  The probability amplitudes   {\it are} well defined quantities in quantum mechanics. They are particular properties of a 
pre- and post-selected quantum system \cite{Feynl}, and we just found one of them to be zero. 
We argue that little else can be added to this conclusion.
The main achievement of the "weak measurement theory" (for a review see \cite{WMrev}) is the discovery of a scheme whereby the response of a weakly perturbed system yields the value of the amplitude in question. Its main problem is not recognising the amplitude for what they are \cite{PLA2016}, \cite{DSEA}.
\section{An amplitude is just an amplitude} 
One may take a view that the only "physical"  quantities are the probabilities and the average values of the observed quantities. The amplitudes, on the other hand, should be just computational tools, no longer needed once the calculation is completed. Is it then surprising that the values of these theoretical constructs can be determined in an experiment?
We argue that it is not. Like the probability itself, a probability amplitude characterises an ensemble. A probability  cannot be measured directly in a single attempt, and requires many trials for the frequency, with which the $i$-th property appears, to approach the probability $P(i)=|A(i)|^2$. But by the time $P(i)$ is evaluated, we will have measured 
indirectly also the modulus of the amplitude $A(i)$, $|A(i)|=\sqrt{P(i)}$. Furthermore, a small perturbation applied to the system 
would change $A(i)$ by $\delta A(i)$, and the probability $P(i)$ by $\delta P(i) \approx 2Re[A(i)\delta A(i)]$. 
Thus, comparing the perturbed and unperturbed frequencies we can indirectly measure the real part 
of a complex valued quantity $A(i)\delta A(i)$. Similarly, in a weak measurement of a projector,
 many accurate observations of the pointer's position, 
 allow us to deduce the value  of $\text{Re} \left [\alpha^{F \gets  I}_i\right ]$.  One's ability to measure probability amplitudes indirectly  is a trivial consequence of the perturbation theory and the structure of quantum probabilities.  However, such measurements
 provide no deeper insight into their physical meaning. The amplitudes remain just something one needs to square in order to arrive at observable frequencies, in other words, the probability amplitudes.
\section{"Weak traces" and interferometers}
In a nutshell, the case of the "three-path interferometer" discussed in \cite{Matz} involves three pathways with 
$A^{F \gets  I}_i \ne 0$, of which two amplitudes have equal magnitudes, but opposite signs, e.g., 
$A^{F \gets  I}_1=- A^{F \gets  I}_2 \ne 0$
(see Fig.1a).
\begin{figure}
	\centering
		\includegraphics[width=8cm,height=6cm]{{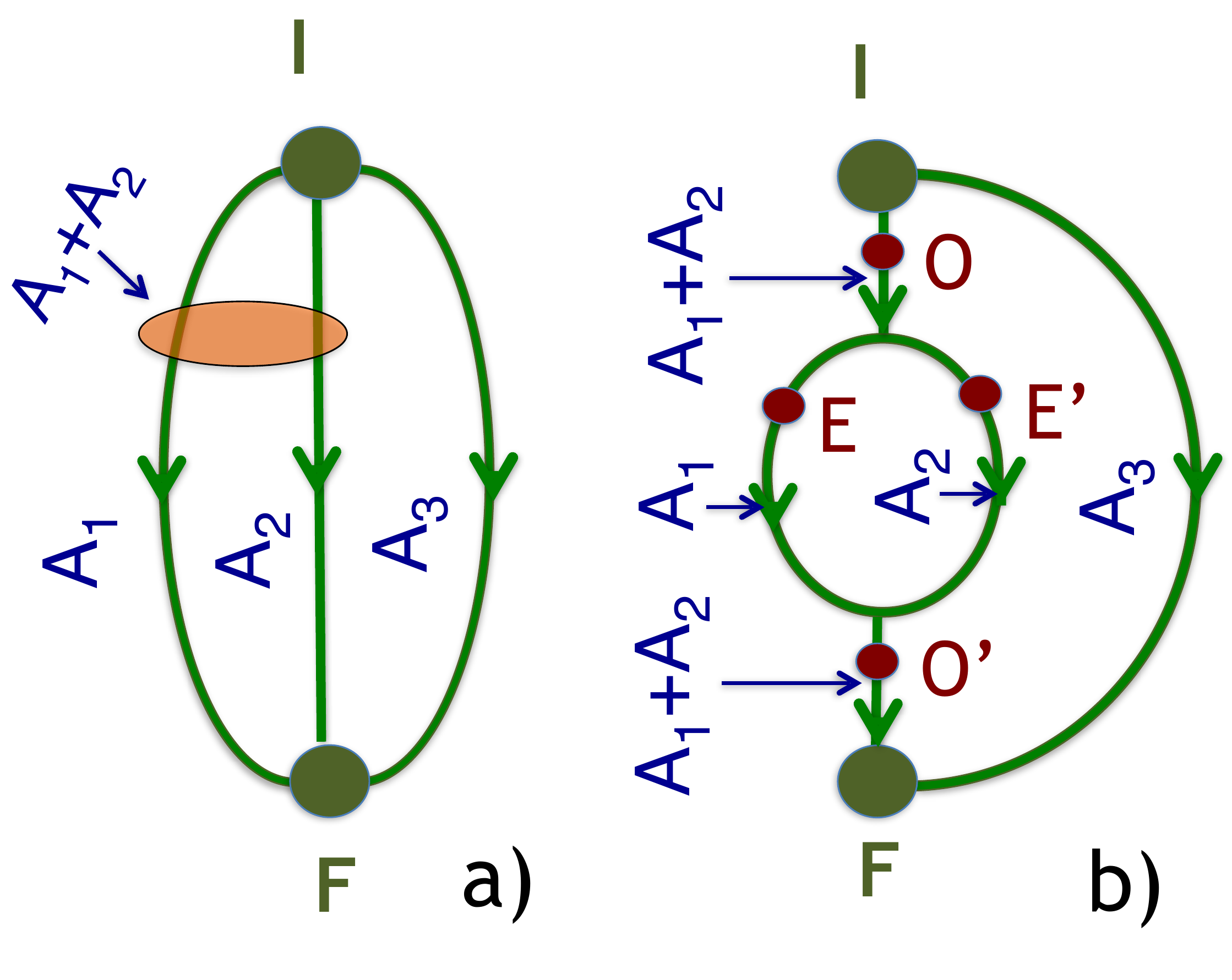}}
\caption{(Color online) 
a) Three virtual paths with relative amplitudes $A_i$, $i=1,2,3$ (we omit the superscript $F\gets I$) connect the initial and the final states of a particle.
 For $A_1=-A_2\ne 0$, and $A_3 \ne 0$, a weak meter would give non-zero readings in the paths $1$ and $2$, but not in their superposition. b) The same system, but with paths $1$ an $2$ merged 
before and after the inner loop. The weak meter would give non-zero readings at $E$ and $E'$, but not at $O$ and $O'$, creating an impression that the particle "was" in the loop ", yet never entered or exited it".
} 
\label{fig:4b}
\end{figure}
Now, from (\ref{3}), weak measurements of operators $\hat {\Pi}_1=|b_1\ra\la b_1|$ or $\hat {\Pi}_2=|b_2\ra\la b_2|$ would detect an non-zero mean shift of the pointer (a weak trace). One is tempted to conclude that the particle "was" wherever the traces were left. But then a measurement
of the projector on the union of the two paths, $\hat {\Pi}_{1+2}=|b_1\ra\la b_1|+|b_2\ra\la b_2|$ would yield no weak trace.
All three measurements can be made simultaneously, e.g., by putting the meters at some locations $O$, $E$,$E'$  and $O'$ (see Fig.1b).
So should we conclude that a quantum particle is in the first and second path, but not in their union (superposition) at the same time \cite{FOOT}? 
\newline
The authors of \cite{Matz} answer the question as follows:
"If the system cannot go through O and be detected in the postselected state, then we can say that 'the particle was not there' provided 'was' is employed in a liberal sense because 'the system' is generally taken to mean 'the system state,' whereas here we are discerning a particular particle property correlated with a transition to a postselected state."
\newline
Our answer is, however, different.
According to (\ref{2}), a pointer placed at $O$ or $O'$ would unite paths $1$ and $2$ in Fig. 1, while leaving path $3$
intact. The amplitude for the combined path $1\cup 2$  is $\tilde A^{F \gets  I}_1 =  A^{F \gets  I}_1+A^{F \gets  I}_2=0$,
and a weak measurement at these two locations would yield a null mean shift  $\text {Re} [(A^{F \gets  I}_1+A^{F \gets  I}_2)/\sum_{j=1}^3A^{F \gets  I}_j]=0$.
A pointer placed at $E$ would unite paths $2$ and $3$, and yield a nonzero mean shift  $\text {Re} [A^{F \gets  I}_1/A^{F \gets  I}_3]\ne 0$. Similarly, a  non-zero pointer shift will be obtained for a pointer placed at $E'$. We have, therefore, obtained the values of the amplitudes for the real pathways that would be created,
should a strong meter be applied at $O$, $O'$, $E$, or $E'$. We argue that the above is the only consistent description offered by conventional quantum mechanics. We refrain from using the notion of a particle "being there" even in the liberal sense employed by the authors of \cite{Matz}, in order to avoid inconsistencies and spurious "paradoxes" that might arise.  
\newline
One such paradox is the case of a photon passing through a "nested Mach-Zehnder interferometer" first studied 
in \cite{Vaid}, and also discussed in  \cite{Matz} and \cite{PLA2017}.
There the photon is thought to be "found" in the nested loop, similar to the one shown in Fig. 1b, but not in the arms leading to and from it  \cite{F1}. 
(We stress that this was not the opinion shared by the authors of \cite{Matz}, who argued instead that the evolving wavefunction physically behaves as
an extended object whose properties relative to pre- and post-selected
states can be measured locally.)
However, accepting the proposition of \cite{F1}, one would need to explain also why the photon is no longer "found" inside the loop if the ingoing arm (where it "never was") is blocked \cite{DSrepl}, so that $A^{F \gets  I}_1=A^{F \gets  I}_2= 0$. If it is proposed that quantum motion 
can be affected by something that happens in the region the particle never visits, a further problem would arise.
An experiment performed in Los Angeles should not depend on what happens in Paris precisely because the particle stays in the lab, and never visits France. Should then there be two senses in which the particle "is not there",
one for a particular part of the interferometer, and another for the rest of the world? It would be prudent to stop here.
 A reasoning which requires ever more assumptions of increasing complexity is unlikely to be helpful, 
and a recourse to the axioms of the theory \cite{Feynl}, however restrictive they may seem, should be preferred \cite{Russ}. 
\section{Null weak value of an arbitrary observable}
Finally, the authors of Ref.\cite{Matz} considered the case of "null weak value of a general observable" where the quantity in the square brackets in Eq.(\ref{4}) (weak value of an operator $\hat{B}$, denoted $B_w$) vanishes for a particular choice of the states $|\psi_F\ra$ and $|\psi_I\ra$. Assuming, for simplicity, that none of the eigenvalues $B_i$ are degenerate, and not all of $ \tilde{A}^{F\gets I}_i$ vanish, we have
\begin{eqnarray}\label{6}
B_w \equiv \sum_k B_k \tilde{\alpha}^{F \gets  I}_m=\frac{\la \psi_F |\hat{U}(t_2,t)\hat{B}\hat{U}(t,t_1)|\psi_I\ra}{\la \psi_F |\hat{U}(t_2,t_1)|\psi_I\ra}\n
=\frac{\la \psi_F (t)|\hat{B}|\psi_I(t)\ra}{\la \psi_F(t)|\psi_I(t)\ra}=0\q\q\q\q\q
\end{eqnarray} 
What could this mean? The authors of \cite{Matz} explain that "A null weak value of an observable A obtained at some location X means that the system property represented by A cannot be found at X and detected in the postselected state."
\newline
Again, we propose a minimalist answer, using only the basic concepts of standard quantum mechanics.
It is instructive to start with a strong measurement of Sect. III. The outcomes of each trial are a value $B_i$ registered by the accurate meter, and a value of $\xi=\pm1$, corresponding to the success ($+1$) or failure ($-1$) of the post-selection in $|\psi_F\ra$. Out of $M$ trials, the combination $(B_i, \xi)$, 
 will occur in $M(B_i,\xi)$ cases.  For many trials, the frequencies of the occurrences will tend to the probabilities in Eq.(\ref{3}),
\begin{eqnarray}\label{7}
\omega_i(M)=M(B_i,1)/M _{M >>1} ^{{\q \longrightarrow}} P^{F\gets I}_i. 
\end{eqnarray} 
Now the average value of $\hat B$ conditional on the successful post-selection, $B_s$, is obtained by writing down the value of $B_i$
each time post-selection succeeds, adding up all values, and dividing the result by the number of entries.
It is the same as the mean pointer shift in Eq.(\ref{3}),
\begin{eqnarray}\label{8}
B_s\equiv \text{lim}_{M\to \infty}\sum_{i=1}^N B_i\omega_i(M)/\sum_{i=1}^N\omega_i(M)=\la f\ra_s
\end{eqnarray}
Our point is this: the result of a series of accurate measurements with post-selection is the probability distribution $P^{F\gets I}_i$. $i=1,2,...N$.
The mean pointer shift $\la f\ra_s$ yields only its first moment.  Finding $\la f\ra_s=B_s=0$ would only indicate the exact cancellation between the terms in the numerator of Eq.(\ref{8}). In other words, we will have proven that the probabilities $P^{F\gets I}_i$ satisfy a particular {\it sum rule},
 $\sum_{i=1}^N B_iP^{F\gets I}_i=0$. It appears that no deeper meaning can be attributed to this result.
 \newline
 The case of a null weak value is not much different. With no probabilities produced, one may only arrive at conclusions above amplitudes. Finding $B_w=0$ would only indicate the exact cancellation between the terms in the numerator of Eq.(\ref{6}),
 and prove a sum rule $\sum_{i=1}^N B_iA^{F\gets I}_i=0$, satisfied by the amplitudes, rather than by the probabilities.
 It is difficult, we argue, to attribute any deeper meaning to this result as well.
To decide whether the interpretation of a null weak value given in \cite{Matz} and cited above, is indeed more informative, 
one requires a detailed explanation of what is meant by the "system property represented by A" \cite{Matz}. 
Without it, we limit ourselves say that a vanishing weak value of a particular operator imposes a restriction on the numerical values 
of the relative amplitudes in Eq.(\ref{6}).
\section{Summary}
In summary, standard quantum mechanics postulates that  a quantum system, making a transition between known initial and final states, is described by transition amplitudes, defined for all possible ways in which the transition can occur \cite{Feynl}.
Quantum "weak values" (WV) are but such amplitudes, or their combinations. Therefore, an attempt to give a "meaning" to the WV amounts to the difficult task of  going beyond the basic axioms of the theory or, in Feynman's words 
"finding the machinery behind the law" \cite{Feynl}. As far as we can see, the "weak measurement theory"  \cite{WMrev} has provided a way of measuring quantum amplitudes, but gained no further insight into their physical significance. In the absence of progress in this direction, any apparently "surprising" or ambiguous statement regarding the behaviour of a pre- and post-selected quantum system can and should be reduced to a statement about the constituent amplitudes, where the enquiry must stop. We argue that this type of analysis will remain the best one available, at least until quantum theory progresses beyond its current fundamental principles.
 \section {Acknowledgements}  Support of
MINECO and the European Regional Development Fund FEDER, through the grant
FIS2015-67161-P (MINECO/FEDER) 
is gratefully acknowledged.

\end{document}